\documentclass[a4paper,twocolumn,showpacs,superscriptaddress,prl]{revtex4}
\usepackage{epsfig}

\begin{document}

\newcommand{\vect}[1]{{\bf #1}}
\newcommand{\comments}[1]{\hfill {\tt {eq:~#1}}}
\newcommand{\dbar}{d\!\!^-}
\title{Current fluctuations near to the 2D superconductor-insulator
  quantum critical point}

\author{A. G. Green}
\address{School of Physics and Astronomy, University of St Andrews,
North Haugh, St Andrews KY16\ 9SS, UK}
\author{J. E. Moore}
\address{Department of Physics, University of California, LeConte
  Hall, Berkeley, CA94720, USA}
\author{S. L. Sondhi}
\address{Department of Physics, Princeton University, Princeton, NJ 08544, USA}
\author{A. Vishwanath}
\address{Department of Physics, University of California, LeConte
  Hall, Berkeley, CA94720, USA}

\date{\today}

\begin{abstract}
Systems near to quantum critical points show universal scaling in their response functions. We consider whether this scaling is reflected in their fluctuations; namely in current-noise. Naive scaling predicts low-temperature Johnson noise crossing over to noise power $\propto E^{z/(z+1)}$ at strong electric fields. We study this crossover in the metallic state at the 2d z=1 superconductor/insulator quantum critical point. Using a Boltzmann-Langevin approach within a $1/N$-expansion, we show that the current noise obeys a scaling form $S_j=T \Phi[T/T_{eff}(E)]$ with $T_{eff} \propto \sqrt{E}$.
We recover Johnson noise in thermal equilibrium and $S_j \propto \sqrt{E}$ at strong electric fields. 
The suppression from free
carrier shot noise is due to strong correlations at the critical
point. We discuss its interpretation
 in terms of a diverging carrier charge
$\propto 1/\sqrt{E}$ or as out-of-equilibrium Johnson noise with
effective temperature $\propto \sqrt{E}$.
\end{abstract}

\maketitle

The notion of quantum criticality provides one of the few general
approaches to the study of strongly correlated quantum many-body
systems\cite{Sachdev1999}. The scale invariance that characterizes the zero
temperature critical point leads to characteristic, universal power-law
dependences for various quantities in its
proximity; these dependences can be computed within a continuum field theory.
A number of theoretical works have calculated the impact of such
universality upon
conductivity\cite{Cha1991,Damle1997,Sachdev1998}. These calculations
provide robust experimental predictions. Two recent works of Dalidovch and
Philips\cite{Dalidovich2004},
and Green and Sondhi\cite{Green05}, have extended the analysis to see whether
universality persists out of equilibrium. Surprisingly it does; at
least in transport.

In this work we consider another experimentally measurable quantity:
current fluctuations. Measurements of current fluctuations are
usually restricted to mesoscopic samples in order that the relative
fluctuations of current be significant.  Mesoscopic samples have
relatively few conducting channels giving a total conductance of
order the conductance quantum $e^2/h$. Quantum-critical systems may
also satisfy this criterion. For example, at the two-dimensional
superconductor-insulator (SI)transition the conductance is also of
order the conductance
quantum\cite{Damle1997,Cha1991,Dalidovich2004,Green05}. 
It would be revealing, therefore, to compare the current fluctuations
at the metallic critical point of bosons with the more familiar
results for non-equilibrium current noise in diffusive Fermi 
liquids\cite{Nagaev}. Transport properties are readily measurable and
theoretical predictions regarding their scaling and universality at
two-dimensional quantum critical points (in charged systems such as SI,
and Quantum Hall Effect transitions) have stimulated many
experiments. Non-equilibrium current
fluctuations may provide an independent window 
upon
these quantum critical points.

There are two contributions to current noise; Johnson noise and Shot
noise. Since it is a reflection of the
fluctuation-dissipation relation, Johnson noise is expected to be
unaffected by correlations between carriers. On the
other hand, shot noise reflects the fact that charge is carried in
quanta. If there are strong correlations between carriers, we expect
it to be significantly altered.

The situation in a non-equilibrium steady state is
trickier; Johnson and shot 
noise
cannot easily be
separated. On the one hand, one may identify an effective
temperature scale $T_{eff}$ for the out-of-equilibrium distribution
and treat the current fluctuations as Johnson-like. Dimensional
analysis suggests $T_{eff} \propto E^{z/(1+z)}$ and current noise
proportional to $E^{z/(1+z)}$. On the other hand, one may
identify the out-of-equilibrium noise with shot noise and a carrier
charge that diverges as $E^{-1/(1+z)}$.

We present a microscopic calculation that recovers these results for
the superfluid to insulator transition of two-dimensional bosons with
particle-hole symmetry. We use a Boltzmann-Langevin\cite{BL1,BL2}
approach within a
$1/N$-expansion\cite{Sachdev1998} to analyse current noise
near this transition.  This ``kinetic theory of fluctuations''
approach is known to give the
correct result for the leading noise correlations in the diffusive
metal, but needs
to be modified for higher moments~\cite{HigherCorrelators}. The system
is considered to be in
thermal contact with a substrate. This, in conjunction with the
internal scattering in the system, leads to a uniform,
non-equilibrium, steady-state distribution, provided
that the length of the sample is greater than the correlation
length\cite{Green05}. This situation is quite different from that
considered elsewhere, where the only coupling to the outside world is
through the leads. It permits the current-noise properties to be entirely
determined by the internal scattering and so to be universal.

Our main results are as follows: in thermal equilibrium, we recover
Johnson noise with noise power $S_j=4\sigma T$
 (as we must in order to satisfy the equilibrium
fluctuations dissipation relation). This crosses over at very large
electric fields to $S_j  \propto \sqrt{E}$. This is strongly
suppressed from the usual shot noise result of $S_j \propto j \propto
E$ due to the strong correlations at the quantum critical point. This
large field result may be considered as either shot noise with a
carrier charge which diverges as $1/\sqrt{E}$ or alternatively as a
non-equilibrium Johnson noise with effective temperature
$T_{eff}= \sqrt{\hbar ceE}\pi^2/4k$. Between these two limits, we expect the
current noise to obey a universal scaling form $S_j=T\Phi[T_{eff}(E)/T]$.

We begin by outlining the field theoretical formulation of the
symmetric, bosonic, superfluid to insulator transition. We then give
Boltzmann and Boltzmann-Langevin equations for the system in thermal
equilibrium. The latter is used to deduce the Johnson
noise. Next, we turn to the Boltzmann and Boltzmann-Langevin equations
for the zero-temperature system under a finite electric field. This is
used to deduce the out-of-equilibrium current-fluctuations. Finally we
turn to a discussion of the implications of these results.

\paragraph*{Field Theory}
%%%%%%%%%%%%%%%%%%%
The critical region of the symmetric superfluid to Mott insulator transition
phase diagram is described by a charged scalar field with a quartic
interaction\cite{Sachdev1999,Damle1997}:
\begin{equation}
{\cal H}
=
\int d^d x
\left[
\Pi^{\dagger} \Pi + \nabla \phi^{\dagger} \nabla \phi +m^2 \phi^{\dagger} \phi
+
\lambda (\phi^{\dagger} \phi)^2
\right],
\label{Hamiltonian}
\end{equation}
where $\phi$ is the complex scalar field and $\Pi$ is its conjugate
momentum. These satisfy the usual commutation relations
$
\left[ \phi({\bf x},t), \Pi({\bf y},t) \right]
=
i \delta({\bf x}-{\bf y})
$.
It is convenient to choose the bare interaction $\lambda$ to have its
fixed point value $u^* \Lambda^{3-d}$ with momentum cutoff $\Lambda$,
\cite{Sachdev1999} although
we will not need the precise, regularization-dependent, value.
At the
zero-temperature critical point, the renormalized mass is zero which
corresponds to a particular choice $m^*$ of the bare mass.
The effects of applying an electric field, ${\bf E}$, are included by
minimally coupling to a vector potential; $\nabla \phi \rightarrow D
\phi=\left(\nabla +i e{\bf A}/\hbar \right)\phi$. We choose the gauge ${\bf
  A}={\bf E}t$. This is equivalent to a contact-free measurement 
 (for example by 
%({\it e.g}  
placing the system on a cylinder and inducing an EMF on its
surface by uniformly increasing the flux through it). We believe that
our results carry over unchanged to the case with good contacts at the ends.

The normal modes of this Hamiltonian (in the absence of interaction)
are charge density fluctuations. These occur with positive and negative
charges, corresponding to decrease or increase in charge density from
the average. We determine the current fluctuations near to the
critical point by considering a Boltzmann equation for the occupation
of these modes. We use $a^{\dagger}$/$b^{\dagger}$ and $a$/$b$ to represent the
creation and annihilation of positively/negatively charged density fluctuations.

{\it The Boltzmann equation} in thermal equilibrium and within a
$1/N$-expansion of 
the Hamiltonian 
Eq.(\ref{Hamiltonian}) is given
by\cite{Damle1997,Measure}
\begin{eqnarray}
& &
\left[ \partial_t +(e{\bf E}/\hbar). \partial_{\bf k} \right]f_{\bf k}
\nonumber\\
&=&
-\int d{\bf q}
\left[
\begin{array}{c}
\gamma_{{\bf k}{\bf q}} f_{\bf k} \left( 1 + f_{\bf q} \right)
-
\gamma_{{\bf q}{\bf k}} f_{\bf q} \left( 1 + f_{\bf k} \right)
\\
\tilde \gamma_{{\bf k}{\bf q}} f_{\bf k} f_{-\bf q}
-
\tilde \gamma_{{\bf q}{\bf k}}
\left(1+ f_{-\bf q} \right)
\left( 1 + f_{\bf k} \right)
\end{array}
\right].
\label{Boltzmann}
\end{eqnarray}
$\gamma_{{\bf k},{\bf q}}$ and $\tilde \gamma_{{\bf k},{\bf q}}$
describe the matrix elements for particle-particle and
particle-anti-particle scattering, respectively\cite{gamma} and
$$
f({\bf k},t)
=
\int d{\bf q}
\langle
a^{\dagger}_{{\bf k}+{\bf q}/2}(t) a_{{\bf k}-{\bf q}/2}(t)
\rangle
$$
denotes the distribution function, with a similar distribution $\tilde
f({\bf k},t)$ for the negatively charged modes. Particle-hole symmetry
implies the relation $f({\bf k},t)= \tilde f(-{\bf k},t)$. We
restrict our explicit consideration to the positively charged channel;
final expressions include the negatively charged channel
through appropriate factors of two.

{\it The Boltzmann-Langevin equation} is an equation describing the
stochastic evolution of fluctuations, $\delta f_{\bf q}({\bf r},t)$, in the
distribution function about its equilibrium or steady-state. In the
present case, it is given by
\begin{eqnarray}
& &
\left[ \partial_t +{\bf v}.\partial_{\bf r} +(e{\bf E}/\hbar). \partial_{\bf q} \right]
\delta f_{\bf k}
=\gamma_{\bf k}
\left[
\delta f_{\bf k}- \delta \bar f_{\bf k}
\right]
+\eta_{\bf q},
\nonumber\\
\nonumber\\
&=&
-\int d{\bf q}
\left[
\gamma_{{\bf k}{\bf q} }
\frac{1+f_{\bf q}}{1+f_{\bf k}}
\delta f_{\bf k}
-
\gamma_{{\bf q}{\bf k} }
\frac{1+f_{\bf k}}{1+f_{\bf q}}
\delta f_{\bf q}
\right]
+\eta_{\bf q}
\nonumber\\
& &
-\int d{\bf q}
\left[
\tilde \gamma_{{\bf k}{\bf q} }
\frac{f_{-\bf q}}{1+f_{\bf k}}
\delta f_{\bf k}
-
\tilde \gamma_{{\bf q}{\bf k} }
\frac{1+f_{\bf k}}{f_{-\bf q}}
\delta f_{-\bf q}
\right]
\label{BoltzmannLangevin}
\end{eqnarray}
where
\begin{eqnarray}
\delta \bar f_{\bf k}
&=&
\int d{\bf q}
M_{{\bf k}{\bf q}} \delta f_{\bf q}
\nonumber\\
M_{{\bf k}{\bf q}}
&=&
\frac{ \gamma_{{\bf q}{\bf k}}   }{\gamma_{\bf k}}
\frac{1+f_{\bf k}}{1+f_{\bf q}}
+
\frac{ \tilde \gamma_{-{\bf q}{\bf k}}   }{\gamma_{\bf k}}
\frac{1+f_{\bf k}}{f_{\bf q}}
\nonumber\\
\gamma_{\bf k}
&=&
\int  d{\bf q}
\left(
\gamma_{{\bf k}{\bf q}}
\frac{1+f_{\bf q}}{1+f_{\bf k}}
+
\tilde \gamma_{{\bf k}{\bf q}}
\frac{f_{-\bf q}}{1+f_{\bf k}}
\right).
\label{overbar}
\end{eqnarray}
The position and time labels of $\delta f_{\bf k}({\bf r},t)$
and $\eta_{\bf q}({\bf r},t)$ have been suppressed for compactness
and the detailed balance conditions
$\gamma_{{\bf q},{\bf k}}f_{\bf q} (1 + f_{\bf k})
=\gamma_{{\bf k},{\bf q}}f_{\bf k} (1 + f_{\bf q})$
and
$\tilde \gamma_{{\bf q},{\bf k}}(1+f_{-\bf q}) (1 + f_{\bf k})
=\tilde \gamma_{{\bf k},{\bf q}}f_{\bf k} f_{-\bf q}$
have been used in order to simplify the right-hand-side.
The time and momentum derivatives on the left-hand-side of
Eq.~(\ref{BoltzmannLangevin})  follow
directly from the Boltzmann equation (\ref{Boltzmann}). The first
term on the right hand side is the linearised form of the scattering
integral. The second is a stochastic term describing fluctuations in
occupation number. It is determined by assuming that the scattering
processes are independently Poisson distributed. Under this
assumption, the quadratic correlations of $\eta_{\bf q}({\bf r},t)$
are given by\cite{ParticleAntiParticle}
\begin{eqnarray}
& & \langle \eta_{\bf q}({\bf r},t) \eta_{{\bf q}'}({\bf r}',t')
\rangle
\nonumber\\
&=&
2\delta_{{\bf r},{\bf r}'}
\delta_{t,t'}
\left[
\begin{array}{c}
(2 \pi)^2\delta_{{\bf q},{\bf q}'}
\int d{\bf k} \gamma_{{\bf q}{\bf k}}
f_{\bf q} (1 + f_{\bf k})
\\
-\gamma_{{\bf q}{\bf q}'}f_{\bf q}(1+f_{{\bf q}'})
\\
+(2 \pi)^2\delta_{{\bf q},{\bf q}'}
\int d{\bf k} \tilde \gamma_{{\bf q}{\bf k}}
f_{\bf q} f_{-{\bf k}}
\\
-\tilde \gamma_{{\bf q}{\bf q}'}f_{\bf q}f_{-{\bf q}'}
\end{array}
\right]
\nonumber\\
&=& 2\delta_{{\bf r},{\bf r}'} \delta_{t,t'}  \gamma_{{\bf q'}}
f_{\bf q} (1 + f_{\bf q})[\delta_{{\bf q},{\bf q}'} -M_{{\bf q}'{\bf
q}}]
%\left[
%\begin{array}{c}
%\delta_{{\bf q},{\bf q}'}
%\gamma_{{\bf q}}
%f_{\bf q} (1 + f_{\bf q})
%\\
%-\gamma_{{\bf q}'}M_{{\bf q}'{\bf q}}f_{\bf q}(1+f_{{\bf q}})
%\end{array}
%\right],
\label{noise}
\end{eqnarray}
where the detailed balance conditions and Eq.~(\ref{overbar})
have been used in order to simplify these expressions. The factor
of two comes from the contributions of in- and out-scattering processes.

{\it Current fluctuations} may be calculated using the
Boltzmann-Langevin equation as follows: first we determine the
local fluctuations in occupation.  In the limit of long times and long
length-scales, we may neglect the derivative
terms on the left-hand-side of Eq.~(\ref{BoltzmannLangevin}). The
fluctuation in the distribution function is then given by
\begin{equation}
\delta f_{\bf k} - \delta \bar f_{\bf k}
=\eta_{\bf k}/\gamma_{\bf k}.
\label{f_solution}
\end{equation}
This integral equation has a formal solution given by
\begin{equation}
\delta f_{\bf k}({\bf x},t) = \int d{\bf q}[{\bf 1}-{\bf
M}]^{-1}_{\bf kq}
\eta_{\bf q}({\bf x},t)/\gamma_{\bf q},
\label{formal_solution}
\end{equation}
where we are using a matrix notation for functions of momentum
with $[{\bf 1}]_{\bf kq}=(2 \pi)^2\delta({\bf k}-{\bf q})$ and the
inverse of a matrix $\bf N$ defined as $\int d{\bf q}[{\bf
N}]^{-1}_{\bf kq}[\bf N]_{\bf qk'}=[{\bf 1}]_{{\bf kk}'}$.

The resulting fluctuation in current is given by:
\begin{equation}
\delta {\bf j} = \int d{\bf p} {\bf v}_{\bf p} \delta f_{\bf q},
\label{currentf}
\end{equation}
where
${\bf v}_{\bf
k}=\partial_{\bf k} \epsilon_{\bf k}$. Using the noise correlations
from Eq.~(\ref{noise}), after some algebra, we obtain the following
expression for the correlation of current fluctuations in thermal
equilibrium \cite{Correlations}:
\begin{eqnarray}
\nonumber 
\langle
 \delta j_{\alpha}({\bf r},t) 
\delta_{\beta}({\bf r}',t')
\rangle 
&=& 
\delta_{\alpha,\beta}\delta_{{\bf r},{\bf r}'}
\delta_{t,t'}
\int d{\bf p} d{\bf q} 
{\bf v}_{\bf p}.
{\bf v}_{\bf q}
\\
& &\times[{\bf 1}-{\bf M}]^{-1}_{{\bf p}{\bf q}}
f_{\bf q} (1 + f_{\bf q})/\gamma_{{\bf q}}
\label{CurrentFlucts1}
\end{eqnarray}
This result should be compared with the result for
fermions\cite{Nagaev} where the Bose enhancement factor $f+1$ is
replaced by a Fermi factor $1-f$. The low frequency current noise
$S_j$ can now be expressed as:
\begin{equation}
S_j=\int ^\infty _{-\infty} dt d{\bf r}\langle {\bf j}({\bf
r},t).{\bf j}(0,0) \rangle
\end{equation}

{\it In Thermal equilibrium}, the current
fluctuations reduce to the Johnson result $S_j=4 \sigma_T T$, with
the universal conductivity 
$\sigma_T$
determined by a linear response expansion of the Boltzmann equation
(\ref{Boltzmann})\cite{LinearResponse}. As
the electric field is increased away from zero, symmetry under ${\bf
E} \rightarrow -{\bf E}$ requires that the lowest order correction
to this result must be proportional to ${\bf E}^2$. A closed form
expression for this correction may be obtained by including the
lowest order, static distortion of the distribution function by the
electric field, $\delta f^s$\cite{LinearResponse}. The resulting
expression is cumbersome and not very revealing. We note, however,
that the corrections to the distribution function are proportional
to powers of $\hbar ceE/(kT)^2$ (to see this, rescale the momentum
integral in the expression for $\delta f^s$ in \cite{LinearResponse}
and use the fact that for critical, relativistic bosons $|{\bf
v}_{\bf
  k}| =c$). The current noise is, therefore, expected to be a
universal scaling function of $\hbar ceE/(kT)^2$;
\begin{equation}
S_j= T f \left[\hbar ceE/(kT)^2 \right] = T \Phi [T_{eff}(E)/T],
\label{scaling_function}
\end{equation}
where we have defined $T_{eff}(E)=\sqrt{\hbar ceE}\pi^2/4k$ (the
reason for the numerical factor
in this expression will become apparent later).
The low-temperature limit of this scaling function is $\Phi(0)= 4
\sigma_T $, where $\sigma_T=0.2154e^2/h$ given by the universal
conductance found in thermal
equilibrium\cite{Sachdev1999,Damle1997}. Next, we turn to a
calculation of the high-field limit of this scaling function.

{\it Out-of-Equilibrium}.
Under the application of a strong electric field, the system is driven
far from thermal equilibrium. In this situation, the Boltzmann equation
(\ref{Boltzmann}) misses important field-induced tunneling processes
(precisely analogous to Zener breakdown or the Schwinger mechanism). A
Boltzmann equation for this situation was previously derived within a
1/N-expansion in Ref.\cite{Green05};
\begin{eqnarray}
\left[ \partial_t +(e{\bf E}/\hbar).\partial_{\bf k}\right]f({\bf k})
&=&
-\Gamma_{\bf k} f({\bf k})
+e^{-\pi \hbar c \epsilon_{{\bf k}}^2/eE}
\delta \left(\frac{\hbar{\bf k.E}}{eE^2}\right),
\nonumber\\
\label{NonLinearBoltzmann}
\end{eqnarray}
where to lowest order in 1/N\cite{Mass}, $\epsilon_{\bf k}^2={\bf k}^2$ and
\begin{eqnarray}
\Gamma_{\bf k}(t)
&=&
\frac{8c}{N}
\frac{  \sqrt{|{\bf k}| + k_{\parallel} }  }{\epsilon_{\bf k}}
\int d{\bf k}'
\frac{ \sqrt{|{\bf k}'| + k'_{\parallel} } }{\epsilon_{{\bf k}'}}
f({\bf k}',t)
\nonumber\\
&=&
\frac{2}{\pi}\frac{1}{\sqrt{N k_{\parallel}}}(eE/\hbar)^{3/4}c^{1/4}.
\label{Gamma}
\end{eqnarray}
Eq.(\ref{NonLinearBoltzmann}) may be integrated to obtain an explicit
solution for the out-of equilibrium distribution
function\cite{Green05, Integrals}.

Following the same procedure as previously, we may derive a
Boltzmann-Langevin equation describing fluctuations in occupation
number. This has a simpler form than in thermal equilibrium since
there is just one process that dominates the contribution to the
fluctuations in occupation number in this strongly
out-of-equilibrium limit. The result is (suppressing position and
time labels)
\begin{eqnarray}
\left[ \partial_t +{\bf v}.\partial_{\bf r}+{\bf E}.\partial_{\bf k}\right]
\delta f_{\bf k}
=
-\Gamma_{\bf k} \delta f_{\bf k} +\eta_{\bf k}
\nonumber\\
\langle
\eta_{\bf q}({\bf r},t) \eta_{{\bf q}'}({\bf r}',t')
\rangle
=
(2 \pi)^2
\delta_{{\bf r},{\bf r}'}
\delta_{t,t'}
\delta({\bf q}-{\bf q}')
\Gamma_{\bf q} f_{\bf q}
\label{NonLinearBoltzmannLangevin}
\end{eqnarray}
Solving, as before, in the limit of long times and large distances, the
fluctuation in occupation number is
given by $\delta f_{\bf q}= \eta_{\bf q}/\Gamma_{\bf q}$
and the correlation of fluctuations in current are given by
\begin{eqnarray}
& &
\langle j_{\alpha}({\bf r},t)j_{\beta}({\bf r}',t') \rangle
\nonumber\\
&=&
2e^2\delta_{\alpha, \beta}
\int d{\bf p} d{\bf q}
\frac{{\bf v}_{\bf p}.{\bf v}_{\bf q} }{\Gamma_{\bf p}\Gamma_{\bf q}}
\langle \eta_{\bf p}({\bf r},t) \eta_{{\bf q}}({\bf r}',t')
\rangle
\nonumber\\
&=&
2e^2 c^2\delta_{\alpha, \beta} \delta_{{\bf r},{\bf r}'}
\delta_{t,t'}
\int d{\bf p}
\frac{f_{\bf p}}{\Gamma_{\bf p}},
\end{eqnarray}
where we have used the fact that ${\bf v}_{\bf p}^2=c^2$ for
relativistic bosons in the large N limit and the factor of two comes
from the contribution of particles and holes. Using the explicit form of
the non-equilibrium distribution function and performing the
integrals, one obtains the high-field contribution to current noise
as
\begin{eqnarray}
\langle j_{\alpha}({\bf r},t)j_{\beta}({\bf r}',t') \rangle
&=&
\delta_{\alpha, \beta}
\delta_{t,t'}
\delta_{{\bf r},{\bf r}'}
\left( \frac{N \pi}{8} \right)^2 e^2 \sqrt{\frac{ecE}{\hbar}}.
\label{HighEnoise}
\end{eqnarray}

 The large field current noise (\ref{HighEnoise})
is dramatically reduced compared with the linear ${\bf E}$-field
dependence of shot noise expected for uncorrelated charge carriers.
It is tempting to ascribe this to an effective charge of carriers
near to the quantum critical point proportional to $1/\sqrt{E}$.
This effective charge shows a dramatic divergence as one approaches
the quantum critical point reflecting the strong correlations of the
system. Whether this picture can be carried through requires an
analysis of higher order statistics of the current
noise\cite{HigherCorrelators}. An alternative, and perhaps more
revealing, interpretation of Eq.~(\ref{HighEnoise}) is as a
non-equilibrium equivalent of Johnson noise. The non-equilibrium
analogue of the fluctuation dissipation relation is recovered if we
identify an effective temperature $T_{eff}= \sqrt{\hbar
ceE}\pi^2/4k$. This is deduced from $4 \sigma_E T_{eff}=2(N\pi/8)^2
\sqrt{ecE/\hbar}$, where $\sigma_E=(N\pi/8) e^2/h$ is the
large-field conductivity\cite{Green05}. The current-noise scaling
function has high-field asymptote $\Phi[x \rightarrow \infty] = 4
\sigma_E x$. To estimate the magnitude of this effect in a physical
setting such as the SI transition in MoGe thin films \cite{Mason},
we can take $E=0.5$V/m, and estimate $c\sim v_F=10^6$m/s. We obtain
$T_{eff}\sim 400$mK, which is an order of magnitude larger than
experimentally accessible temperatures. 
The current noise will, therefore, be enhanced over its thermal value.

\vspace{0.1in}
\noindent
{\it In conclusion}, we have considered the current noise at the
2-dimensional, z=1 superconductor to insulator quantum critical
point. We find that this noise follows a universal scaling function,
$
S_j
= T \Phi[T_{eff}(E)/T],
$
with $T_{eff}=\sqrt{\hbar ceE}\pi^2/4$. This scaling function recovers
Johnson noise in thermal equilibrium and crosses over to an unusual
$\sqrt{E}$-dependent non-linear shot noise or non-equilibrium Johnson
noise at high fields. This is a particular case of a more general
$E^{z/(1+z)}$ scaling expected for high-field current noise. In this
way, current-noise may reveal the universal non-linear scaling
exponents predicted near to quantum phase transitions.

{\sc Acknowledgments:} This work was completed with the support of
the Royal Society, NSF DMR-0238760 and DMR-0213706, the LDRD program of LBNL under
DOE grant DE-AC02-05CH11231 and the A. P. Sloan Foundation (A.V.). We would like to thank D. Gutman and N. Mason for discussions.


\begin{thebibliography}{99}
\bibitem{Correlations}
In matrix notation the equation (\ref{noise}) reads:
$$
\langle \eta_{\bf q}({\bf r},t) \eta_{{\bf q}'}({\bf r}',t')\rangle
= 2\delta_{{\bf r},{\bf r}'} \delta_{t,t'}  \gamma_{{\bf q'}} f_{\bf
q} (1 + f_{\bf q})[{\bf 1}-{\bf M}]_{{\bf q}'{\bf q}}
$$
Combining this with Eq.~(\ref{formal_solution}) we find the
%Combining Eq.(\ref{noise}) with Eq.~(\ref{formal_solution}) we find the
fluctuations of the distribution function are correlated as:
$$
\langle \delta f_{\bf p}({\bf r},t) \delta f_{\bf q}({\bf
r}',t')\rangle = 2\delta_{{\bf r},{\bf r}'} \delta_{t,t'}[{\bf
1}-{\bf M}]^{-1}_{{\bf p}{\bf q}} \frac{f_{\bf q} (1 + f_{\bf
q})}{\gamma_{{\bf q}}}
$$
hence correlators of the current fluctuations (\ref{currentf}) are:
$$
\langle \delta j_{\alpha}({\bf r},t) \delta_{\beta}({\bf
r}',t')\rangle = \delta_{\alpha,\beta}\delta_{{\bf r},{\bf r}'}
\delta_{t,t'}\int_{\bf pq}[{\bf 1}-{\bf M}]^{-1}_{{\bf p}{\bf q}}
f_{\bf q} (1 + f_{\bf q})/\gamma_{{\bf q}}
$$
\bibitem{Mass} Within the 1/N-expansion for the out-of-equilibrium
  system, the gap 
  %$m \propto 1/N$. It 
  does not contribute to the
  current response at lowest order in $1/N$ and is neglected here 
  (See also \cite{Green05}).
\bibitem{Integrals}
The explicit solution for the distribution function is
$
f_{\bf k}
=
\theta(K_{\parallel})
\exp \left[
-\pi K_{\perp}^2 - \gamma \sqrt{ 2 K_{\parallel}}
\right],
$
where the momenta have been rescaled according to ${\bf K}={\bf
  k}/\sqrt{eE/\hbar c}$ and $\gamma$ is determined by
$
\Gamma_{\bf k}
=
\gamma (eE/\hbar)^{3/4} c^{1/4}/\sqrt{2 k_{\parallel}}.
$
In terms of these  variables,
Eq.~(\ref{NonLinearBoltzmann}) may be integrated to obtain the above
solution. The integrals in Eq.~(\ref{Gamma}) may be carried out noting
that $k_{\parallel} \gg k_{\perp}$ to obtain $\gamma^2=8/N\pi^2$.
\end{thebibliography}
\end{document}